\documentclass[journal]{IEEEtran}

\usepackage{comment}

\ifCLASSINFOpdf
   \usepackage[pdftex]{graphicx}
   \usepackage{amsmath,siunitx}
   \usepackage{multirow}
   \usepackage{float}
    \usepackage{placeins}
    \usepackage{color,soul}
\makeatletter
\newcount\SOUL@minus 
\makeatother

\else

   \usepackage[dvips]{graphicx}
   \graphicspath{{IEEEtran/}}
   \DeclareGraphicsExtensions{.eps}
\fi

\usepackage{amsmath}

\begin{document}

\title{A Composite T60 Regression and Classification Approach for Speech Dereverberation}

\author{Yuying~Li,~\IEEEmembership{Student Member,~IEEE,}
        Yuchen~Liu,~\IEEEmembership{Student Member,~IEEE,}
        and~Donald~S.~Williamson,~\IEEEmembership{Senior Member,~IEEE}
\thanks{Y. Li is with the Department
of Intelligent Systems Engineering, Luddy School of Informatics Computing and Engineering, Indiana University, Bloomington,
IN 47408 USA (e-mail: liyuy@iu.edu).}
\thanks{Y. Liu is with the Department of Computer Science, Luddy School of Informatics Computing and Engineering, Indiana University, Bloomington, IN 47408 USA (e-mail: liu477@iu.edu).}
\thanks{D. S. Williamson is with the Department of Computer Science and Engineering, The Ohio State University, Columbus, OH 43065 USA (e-mail: williamson.413@osu.edu).}
\thanks{This work has been submitted to the IEEE for possible publication. Copyright may be transferred without notice, after which this version may no longer be accessible.}}

\maketitle

\begin{abstract}
 Dereverberation is often performed directly on the reverberant audio signal, without knowledge of the acoustic environment. Reverberation time, $T_{60}$, however, is an essential acoustic factor that reflects how reverberation may impact a signal. In this work, we propose to perform dereverberation while leveraging key acoustic information from the environment. More specifically, we develop a joint learning approach that uses a composite $T_{60}$ module and a separate dereverberation module to simultaneously perform reverberation time estimation and dereverberation. The reverberation time module provides key features to the dereverberation module during fine tuning. We evaluate our approach in simulated and real environments, and compare against several approaches. The results show that this composite framework improves performance in 
 environments.
\end{abstract}


\IEEEpeerreviewmaketitle

\section{Introduction}

\IEEEPARstart{R}{everberation} occurs in everyday environments, due to the reflection of sounds off the many surfaces in a room, such as the furniture, walls and floors. This causes listeners to hear a combination of the direct speech signal and the reflections. The effects of reverberation often smear speech across time and frequency, which negatively impacts individuals with impaired hearing \cite{nabelek1974monaural,payton1994intelligibility}, since reverberation degrades perceptual quality and intelligibility. This also creates challenges to various voice-based applications, including automatic speech recognition (ASR) \cite{kinoshita2013reverb, giri2015improving}, speaker identification \cite{zhao2014robust, akula2009speaker} and speaker localization \cite{chakrabarty2017broadband, akula2008compensation}, to name a few.

Current monaural approaches often use deep neural networks (DNN) to remove reverberation. A spectral mapping method \cite{han2015learning}, proposed by Han \textit{et al.}, maps the noisy and reverberant signal to an anechoic signal in the time-frequency (T-F) domain using a fully-connected DNN. It has a post-processing stage that performs iterative phase reconstruction to re-synthesize the estimated time-domain signal. Xiong \textit{et al.} {\cite{mlp}} use a multi-layer perception (MLP) to estimate T60 using features from a Gabor filterbank. A DNN estimates the complex ideal ratio mask (cIRM) \cite{williamson2017time}, which processes the magnitude and phase responses in the imaginary and real domains. The approach simultaneously handles noisy and reverberant conditions. The usage of DNNs, however, is a limitation since they do not capture long-term contextual information. To overcome this limitation, Santos \textit{et al.} proposed a dereverberation method that uses a recurrent neural network (RNN) \cite{santos2018speech} to capture long-term contextual information, along with employing a 2-D convolutional encoder to extract local contextual features. Another RNN model with a long short-term memory (LSTM) network \cite{zhao2018late} proposed by Zhao \textit{et al.} predicts late reflections and subtracts them from the reverberant signal to estimate the direct and early components of reverberation. Zhao \textit{et al.} later propose a dereverberation model that uses temporal convolutional networks (TCN) with a self-attention module \cite{zhao2020monaural}. This method uses self attention to extract dynamic features from the input, and uses the TCN to learn the non-linear mapping from reverberant to anechoic speech. All the above approaches operate in the T-F domain.

Traditional signal processing methods have also been developed. These approaches can be divided into two categories: spectral subtraction and inverse filtering. Lebart \textit{et al.} proposed a spectral subtraction approach \cite{lebart2001new} that removed reverberation by canceling the smearing effects in phonemic energy using prior knowledge of the reverberation time and phonemes. More specifically, the approach estimated the power spectral density (PSD) of the reverberation. The square root of the estimated PSD is subtracted from the reverberant signal, resulting in the estimation of the dereverberated signal's spectrum. Yoshioka \textit{et al.} provided a generalized subband-domain multi-channel linear prediction approach (also known as weighted prediction error-WPE) without prior knowledge of acoustic conditions \cite{yoshioka2012generalization}. Tomohiro \textit{et al.} {\cite{nakatani2010speech} used  a delayed linear prediction (DLP) model to cancel the late reverberation without prior knowledge of the room impulse responses (RIR). WPE estimates a filter that predicts the reverberation tail and subtracts it from the reverberant signal to obtain the maximum likelihood estimate. It has been used in many applications \cite{delcroix2014linear, yoshioka2015environmentally}.

\begin{figure*}[!htb]

\minipage{0.333\textwidth}
  \includegraphics[width=\linewidth]{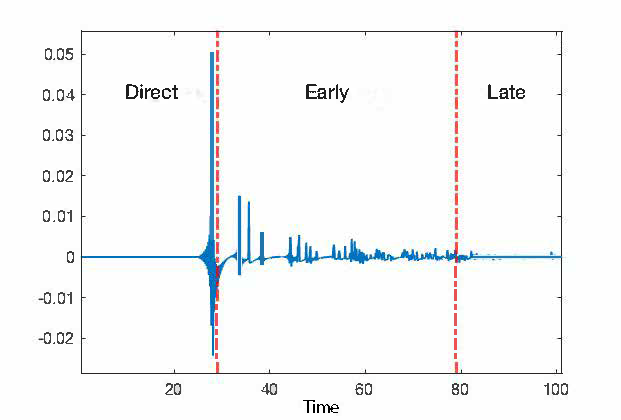}
\endminipage\hfill
\minipage{0.333\textwidth}
  \includegraphics[width=\linewidth]{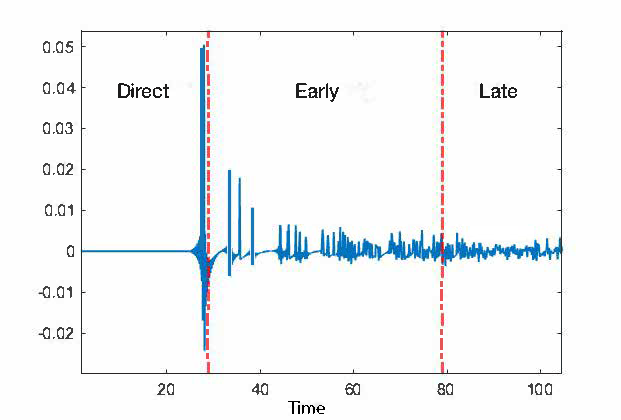}
\endminipage\hfill
\minipage{0.333\textwidth}%
  \includegraphics[width=\linewidth]{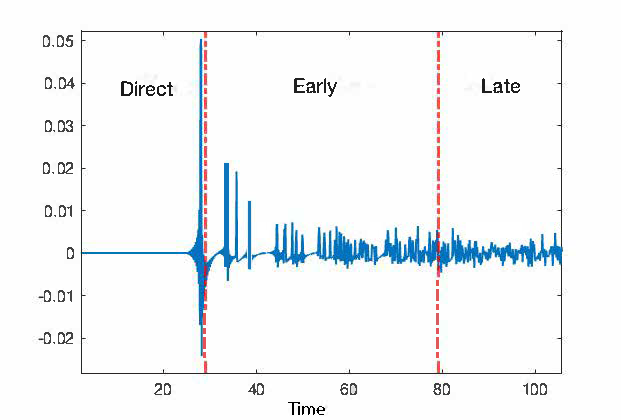}
\endminipage
\caption{(Color Online) The decomposition of a RIR into its direct path, early reflections and late reverberations. From left to right: $T_{60} = 0.3$s, $0.6$s, $0.9$s.}
\label{fig:1}
\end{figure*}
All the above approaches operate directly on the reverberant signal and are either (1) agnostic of the acoustical and contextual information about the room and signal or (2) they assume this information is known and do not estimate it. In particular, the approaches do not leverage or estimate information about the reverberation time, $T_{60}$, which is a strong indicator of the smearing effects of reverberation. It is possible to estimate this information, for instance Bryan \textit{et al.} {\cite{bryan2020impulse}} use a convolutional neural network (CNN) to downsample the input to a single $T_{60}$ value. Considering how important the room environment is to dereverberation, Wu \textit{et al.} \cite{wu2016reverberation} investigate how different context information affect the suppression of reverberation. The approach uses a reverberation-time-aware DNN  that estimates the reverberation time based on the proper selection of the frame shift and context window sizes during feature extraction. It then supplies the log-power spectrogram to the DNNs for dereverberation. Instead of manually generating different contextual information based on the selected frame shift and window size, a self-attention module learns the different representations automatically. This method, however, requires manually choosing the contextual information for different $T_{60}$s and a reliable $T_{60}$ estimator. This reverberation-time-aware DNN has been used as a front-end process for ASR \cite{wu2017end}. Another temporal-contextually aware approach is proposed by Wang \textit{et al.}~\cite{wang2021tecanet}. The main dereverbation model uses time-aware context frames to predict the dereverbation spectrogram, while jointly optimizing the reverberation time. This environment-aware network jointly performs reverberation-time estimation and dereverberation, however, the approach does not always generalize and needs improved performance in real-world settings. These approaches indicate that optimizing with reverberation time offers benefits to dereverberation.

In this paper, we propose a joint-learning approach for speech dereverberation that accurately estimates $T_{60}$ and late reflections.  Early reflections are beneficial to speech intelligibility, so we decided to remove the late reflections \cite{bradley2003importance, hu2014effects}. We separately train a $T_{60}$ estimator that matches the model from our preliminary work \cite{li2021}, and a dereverberation network \cite{zhao2018late}. 
Additional features from the $T_{60}$ estimator are also provided to the dereverberation module. The new feature connects the two networks, and fine-tuning is subsequently performed to generate the final dereverberated result. We do run experiments to determine the loss function from \cite{li2021} that is best for the joint approach. There are four major differences between the adopted dereverberation model and the model in \cite{zhao2018late}: 1) the window and FFT sizes, 2) we stacked three LSTM layers instead of two, 3) we remove the weight-dropping approach during training, but instead use dropout and train with a larger and more balanced dataset. Finally, 4) we also incorporate different input features. 

The rest of the paper is organized as follows. Background information is provided in section \ref{alg}. A detailed algorithm description is provided in section \ref{ad}. The experiments are explained in section \ref{sec:exp}. The evaluation of the experiments and the results are provided in section \ref{sec:eva}. A discussion of related issues and conclusions are mentioned in sections \ref{sec:con} and \ref{sec:newcon}.

\section{Background}
\label{alg}

A reverberant signal, $x(t)$, can be modeled as the convolution of an anechoic speech signal, $s(t)$, with a RIR, $h(t)$ \cite{oppenheim1999discrete}, $x(t)=s(t) * h(t)$, where $*$ denotes convolution, and $t$ denotes the time index. The RIR can be decomposed into three parts: the direct RIR, early RIR and late RIR: $h(t)=h_{d}(t)+h_{e}(t)+h_{l}(t)$, where $h_{d}(t)$, $h_{e}(t)$, $h_{l}(t)$ correspond to direct, early and late components, respectively. Example decompositions are shown in Fig. \ref{fig:1}. The direct component $h_{d}(t)$ starts at the beginning of the RIR and ends approximately 1ms after the first peak, while the early component  extends roughly 50 ms after the direct sound \cite{bradley2003importance}. The late component starts from the end of $h_{e}(t)$ to the end of the RIR. The transition time between the early and late reflections should be between 50 and 150 ms \cite{hidaka2007new}, where a value in this range has often been used \cite{hu2014effects, habets2008temporal, gul2022recycling}. Here, we fix the transition time between the early and late reflections to 50 ms. Each RIR component has the same length as $h(t)$, but they are zero valued outside of the above mentioned intervals. 

Using the distributive property, a reverberant signal can be modeled as the sum of the three RIR components that are convolved with an anechoic speech signal, resulting in the direct sound, $x_{d}(t)$, early reflections, $x_{e}(t)$, and late reflections, $x_{l}(t)$, as shown in the below equation.
\begin{equation}
\label{sig2}
\begin{aligned}
x(t)&=s(t) * (h_{d}(t)+h_{e}(t)+h_{l}(t)) \\
&=s(t) * h_{d}(t) + s(t) * h_{e}(t) + s(t) * h_{l}(t) \\
&=x_{d}(t) + x_{e}(t) + x_{l}(t) \\
&= x_{de}(t) + x_{l}(t)
\end{aligned}
\end{equation}
$x_{de}(t)$ denotes the direct-early component that is defined as the summation of the direct sound and early reflections.

The reverberation time, $T_{60}$, is the time required for the sound in a room to decay 60 dB. It can be modeled using the Sabine formula \cite{young1959sabine}:
\begin{equation}
\label{T60}
T_{60}=0.16V \slash \alpha S 
\end{equation}
where $V$ is the volume of the room, $S$ is the area of its surfaces, and $\alpha$ denotes the absorption coefficient. Longer reverberation times indicate that more reflections occur, which leads to more smearing across time and frequency. Hence the reverberation time has a functional relationship with the RIR. This is also depicted in Fig. \ref{fig:1}, which shows the RIRs generated using different reverberation times. 

\begin{figure*}[!htb]
\begin{center}
\includegraphics[width=\textwidth,height=8cm]{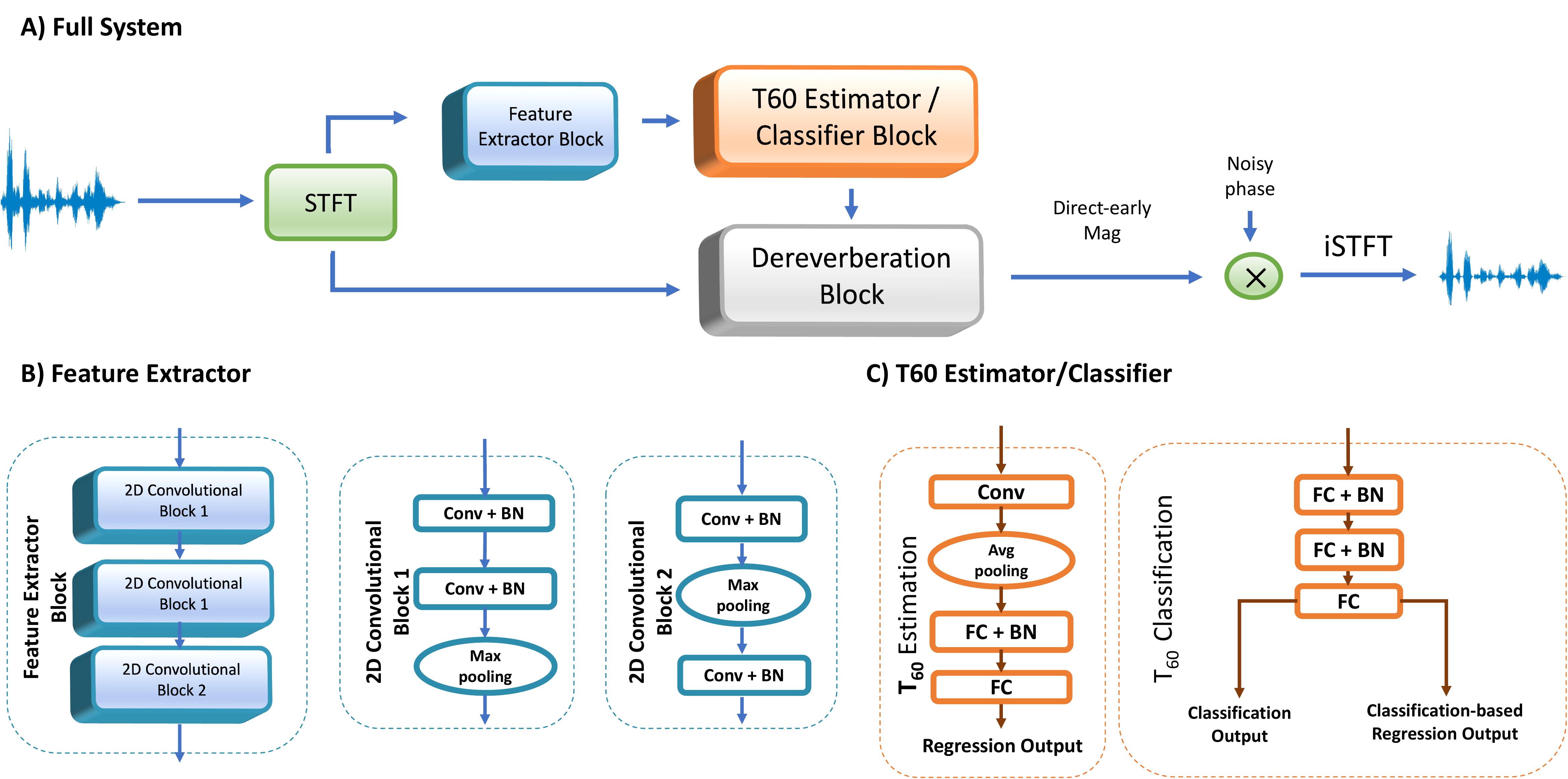}
  \caption{(Color Online) The proposed joint network is shown, where (A) shows the complete approach (B) displays the feature extraction module and (C) depicts the network architecture for the T$_{60}$ classifier and estimator. The dereverberation block is illustrated in Fig. \ref{fig:3}. }
\vspace{-2em}
  \label{fig:2}
\end{center}
\end{figure*}

\section{Algorithm Description}
\label{ad}

We propose to use a DNN to jointly estimate the reverberation time, $T_{60}$, and the direct-early component of reverberant speech. Our proposed approach consists of three stages: (1) $T_{60}$ estimation, (2) direct-early component estimation, then (3) jointly perform  $T_{60}$ estimation and dereverberation. A depiction of our approach is shown in Fig. \ref{fig:2}. 

\subsection{$T_{60}$ Estimation and Classification}
\label{t60}

We treat $T_{60}$ estimation as a multi-task problem, where we simultaneously (a) estimate and (b) classify the reverberation time directly from the reverberant signal.  We adopt this approach because multi-task learning has shown to be beneficial for speech enhancement \cite{eskimez2021human,peng2021attention, chen2015speech}, and when estimating perceptual quality metrics \cite{dong2019classification, zhang2021end}. 

\subsubsection{Features}
We compute the short-time Fourier transform (STFT) of the reverberant signal, where then the log-magnitude response is concatenated with the $sin$ and $cos $ of the phase, $\theta$, across all T-F bins. This is done because prior work has shown that including phase information helps improve performance \cite{wang2018end}. We normalize this concatenated feature so that it is zero mean and unit variance at each frequency bin, across all the training samples. 
\subsubsection{Network architecture}
The normalized input is given to a feature extraction module, which is shown in Figure~\ref{fig:2}(B). It consists of six convolutional layers (Conv), where each Conv layer uses batch normalization (BN) during training. Max pooling is used after every second Conv layer to downsample the latent representation, except for the latter two layers where max pooling is inserted in-between the last two Conv layers. %
This architecture performed well on our preliminary experiments~\cite{li2021} when we evaluated different cost functions and it also has shown to perform well on related speech processing tasks \cite{wang2018multiobjective, xu2018shifted}.

Figure \ref{fig:2}(C) depicts the network architecture that jointly performs $T_{60}$ regression and classification. The latent representations that are generated from the feature extractor serve as the input to the composite $T_{60}$ estimation stage. The composite $T_{60}$ estimation stage consists of two branches, each of which is supplied the same input: (1) the regression only branch (e.g., the left branch), and (2) the combined classification and classification-based regression branch (e.g., the right branch). A similar architecture was proposed in \cite{wang2018study, li2019improved}. For the regression only branch, the shared input is supplied to a Conv layer with rectified linear (ReLU) activation function. This is followed by an average pooling layer, a fully connected layer with batch normalization and a leaky ReLU activation, and a fully connected layer that generates an estimated $T_{60}$ value. 
Figure \ref{fig:2}(C)(right) shows the network details for the $T_{60}$ classification portion of the composite estimation approach. It consists of two FC layers as hidden layers, where each one is followed by a BN layer. A leaky ReLU activation function is used in both FC layers. After the hidden layers, a linear layer serves as an output layer, where the output is further split into two sub portions: $T_{60}$ classification and $T_{60}$ classification-based regression. The former sub-portion uses a softmax activation and is represented by the classification output in Figure \ref{fig:2}(C).  
The latter is based on the following regression loss function:
\begin{equation}
C_{Reg\_T_{60}}=\sum_{i=1}^{H}C_{\text {out }}^{i} \times T_{i}
\end{equation}
where $\times$ is element-wise multiplication, H denotes the number of classes, $T_{i}$ denotes the $T_{60}$ time of the $i$-th class, $C_{\text {out }}^{i}$ is the estimated probability for the $i$-th class.

\subsection{Direct-early Component Estimation}
\label{early}
\begin{figure}
\begin{center}
\includegraphics[width=\columnwidth,height=6cm]{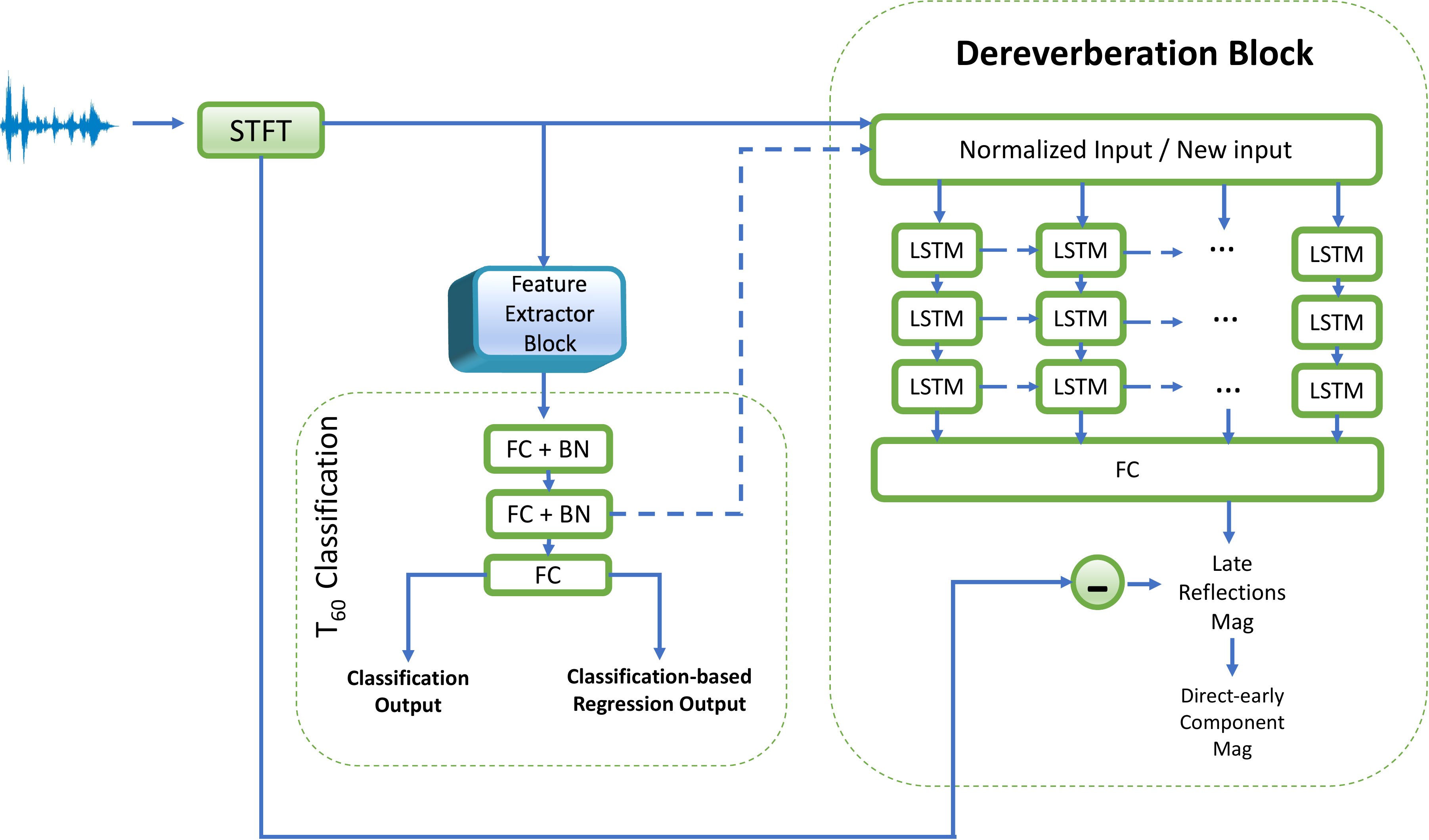}
  \caption{Illustration of our proposed dereverberation block. The dashed arrow only appears during fine tuning.}
\vspace{-2em}
  \label{fig:3}
\end{center}
\end{figure}
Figure \ref{fig:3} depicts our proposed dereverberation process. Prior work shows that spectral subtraction for dereverberation can be performed in the T-F domain for removing late reflections \cite{boll1979suppression, zhao2018late}, so we elect to perform spectral subtraction after the deep neural network estimation. 

\subsubsection{Features and training target}

We first compute the STFT, and use the cubic-root compressed magnitude as the input feature and the cubic-root magnitude of the direct-early component as the training target for the dereverberation module,
which is similar to \cite{zhao2018late}.

\subsubsection{Network Architecture}

The compressed-magnitude features are supplied to a unidirectional LSTM, since they learn long-term dependencies.
The LSTM network is shown in Figure \ref{fig:3} within the dereverberation block. It consists of 3 LSTM layers. A FC layer follows that maps the hidden states to the output, where we predict the late reflections of the compressed magnitude response. We then perform spectral subtraction by subtracting the estimated late reflections from the cubic root magnitude of the corresponding reverberant signal. This results in an estimate of the direct-early component.  

\subsection{Joint Network}
\label{joint}

The two networks shown in Figures \ref{fig:2} and \ref{fig:3} are initially trained separately. In a final step, we propose to combine these two networks and finetune training to form a new joint network, as shown in Figure \ref{fig:3}. During joint training, the output from the penultimate layer in the $T_{60}$ classification block is used as part of the input to the dereverberation block, where it is concatenated with the compressed magnitude.  This forms a newer version of input features that contains both the audio features and reverberation time features. We experimented with different inputs that connected the two modules: 1) the scalar output from the regression-only branch, 2) the one-hot vector from the classification branch, and 3) the penultimate output. We use the penultimate output of the $T_{60}$ estimator, since it is more likely to contain relevant information about the reverberation time \cite{chi2021audio} and performed better empirically.

\subsection{Cost Functions}

The proposed cost function for pre-training the $T_{60}$ branch is shown below, where it is based on our prior work \cite{li2021}:
\begin{equation}
\label{cost1}
\begin{aligned}
L^{A} &= \beta \times \left(\alpha \times L_{\text {cls}}+(1-\alpha) \times L_{\text {creg}}\right)+
(1-\beta) \times L_{\text {reg }}\\
&-\left|\rho_{\text {reg}}\right|-\left|\eta_{\text {reg}}\right|-\left|\rho_{\text {cls}}\right|-\left|\eta_{\text {cls}}\right|
\end{aligned}
\end{equation}
where $L^{A}$ is the loss combination of the cross-entropy, $L_{\text {cls}}$, and mean-square error (MSE) loss terms, $L_{\text {creg }}$ and $L_{\text {reg}}$, for the classification and regression branches, respectively. $\beta$ controls the weight between the losses from the two branches. $\alpha$ balances the cross-entropy loss and the MSE of the classification-based regression task in the classification branch. $L_{\text {creg }}$ minimizes the MSE between the estimated classification-based $T_{60}$ and the ground truth $T_{60}$. This composite loss also includes terms based on Pearson’s correlation coefficient (PCC, $\rho$) and Spearman’s rank correlation coefficient (SRCC, $\eta$), which are normally used as evaluation metrics. $|\cdot|$ denotes the absolute value.

The proposed cost function for the joint network is: 
\begin{equation}
\label{cost2}
L_{\text {joint }}^{B} = \gamma *[\alpha \times L_{c l s}+(1-\alpha) \times L_{\text {creg }}]+(1-\gamma)L_{\text {derev}}
\end{equation}
where $\gamma \in[0,1]$ controls the weight of two parts of the network, $\alpha$ weighs the $T_{60}$ estimation branches, and $L_{\text {derev}}$ denotes the MSE loss calculated in the dereverberation branch. From our prior work on $T_{60}$ estimation, the classification branch outperformed the regression branch, hence we use the classification branch's loss and the dereverberation loss to update the joint network during fune tuning.

\section{Experiments}
\label{sec:exp}


\subsection{Data}

We train with the TIMIT corpus \cite{TIMIT}, which has been used in prior studies \cite{togami2020joint, feng2021dnn}, where it contains 630 native English speakers from eight regions of the United States. We randomly select 5000, 500 and 500 signals to construct our training, validation and testing datasets in simulated environments. All 6000 signals are downsampled to 8 kHz. We simulate RIRs for 14 different room dimensions by using the imaging method \cite{rir}. The dimensions for each room are listed in Table \ref{sim_RIR}. Rooms 1 through 10 are used to generate training and development, while room 11 through 14 are used to generate the testing set
. The distance between the microphone and speaker is set to  1 m across all cases to stabilize the direct to reverberation ratio (DRR) for each condition. We select thirteen reverberation times: 0.3 s to 1.5 s, with increments of 0.1 s to assess different levels of reverberation. We use all thirteen reverberation times to pre-train the $T_{60}$ module, while we use three reverberation times: 0.3 s, 0.6 s and 0.9 s to pre-train the dereverberation block, which we adopted from previous study settings \cite{han2015learning, zhao2018late}. We use the same three reverberation times when fine-tuning the joint network. Including more reverberation times to train the $T_{60}$ module helps with generalization. 
\begin{table}[tbh!]

\centering
\caption{Room dimensions for rectangular simulated RIRs}
\label{sim_RIR}
{\begin{tabular}{c|c|c}
\hline
Group & Name & Dim. (meter)                   \\ 
\hline
\multirow{10}{*}{Seen Rooms} & Room 1 & 9 $\times$ 8 $\times$ 7                    \\ 
                             & Room 2 & 10 $\times$ 7 $\times$ 3                    \\ 
                             & Room 3 & 6 $\times$ 6 $\times$ 10                    \\ 
                             & Room 4 & 8 $\times$ 10 $\times$ 4                    \\ 
                             & Room 5 & 7 $\times$ 7 $\times$ 8                     \\ 
                             & Room 6 & 7 $\times$ 9 $\times$ 5                     \\ 
                             & Room 7 & 8 $\times$ 8 $\times$ 10                    \\ 
                             & Room 8 & 10 $\times$ 10 $\times$ 8                   \\ 
                             & Room 9 & 8 $\times$ 8 $\times$ 6                     \\ 
                             & Room 10 & 7 $\times$ 8 $\times$ 6                    \\ \hline
\multirow{4}{*}{Unseen Rooms} & Room 11 & 9 $\times$ 9 $\times$ 10                   \\
                              & Room 12 & 9 $\times$ 7 $\times$ 9                    \\
                              & Room 13 & 9 $\times$ 10 $\times$ 5                   \\
                              & Room 14 & 10 $\times$ 10 $\times$ 7                  \\ \hline
\end{tabular}}
\end{table} 
\begin{table}[]

\centering
\caption{Room characteristics for Real RIRs: ACE corpus}
\label{real_rir}
{\begin{tabular}{c|c|c}
\hline
Name           & Dim. (meter)     & $T_{60}$(s) \\ \hline
Office 1       & 4.8 $\times$ 3.3 $\times$ 3.0  & 0.34    \\ 
Office 2       & 5.1 $\times$ 3.2 $\times$ 2.9  & 0.39    \\ 
Meeting Room 1 & 6.6 $\times$ 4.7 $\times$ 3.0  & 0.44    \\ 
Meeting Room 2 & 10.3 $\times$ 9.2 $\times$ 2.6 & 0.37    \\ 
Lecture Room 1 & 6.9 $\times$ 9.7 $\times$ 3.0  & 0.64    \\ 
Lecture Room 2 & 13.4 $\times$ 9.2 $\times$ 2.9 & 1.25    \\ 
Building Lobby & 5.1 $\times$ 4.5 $\times$ 3.2  & 0.65    \\ \hline
\end{tabular}}
\end{table}
We simulate 500 different RIRs for each $T_{60}$ in the first 10 room settings, which are used to generate the training dataset. Another 50 RIRs are simulated for each $T_{60}$ for validation. As a result, we have 10 $\times$ 50 = 500 RIRs for each $T_{60}$ in total for the validation set
. For the testing rooms, we use 500 RIRs for each $T_{60}$ and each room setting. As a result, we have 4 $\times$ 500 = 2000 RIRs for each $T_{60}$ in the testing set. We convolve each RIR with one unique speech signal for each $T_{60}$. As a result, during pre-training for $T_{60}$ estimation, we have 5000 $\times$ 13 = 65000 reverberant signals in the training set and 500 $\times$ 13 = 6500 reverberant signals in the validation set. 
During the pre-training stage of the dereverberation block and the joint-learning stage, we have 5000 $\times$ 3 = 15000 reverberant signals in the training set and 500 $\times$ 3 = 1500 reverberant signals in the validation
and 2000 $\times$ 3 = 6000 in the testing set. We further generate unseen non-rectangular rooms using Pyroomacoustics \cite{scheibler2018pyroomacoustics} image source model/ray tracing (ISM/RT) simulator \cite{vorlander2020auralization, schroder2011physically}. We simulate 3 L-shaped rooms with respective dimensions of: 8.5 m $\times$ 3 m $\times$ 6 m and 2 m $\times$ 4 m $\times$ 6 m, 10 m $\times$ 5 m $\times$ 10 m and 3 m $\times$ 1.5 m $\times$ 10 m, and 7 m $\times$ 4 m $\times$ 10 m and 1.5 m $\times$ 2 m $\times$ 10 m. The distance between the source and the speaker is roughly 1.88 m. We simulate 200 RIRs for each room, which results in a total of $200 \times 3 = 600$ RIRs. We convolve each of the RIRs with one signal from the TIMIT test set, resulting in a total of 600 signals. The {$T_{60}$}s range from 0.5 s to 1 s. 
We zero pad the RIRs to match the longest RIR in the dataset \cite{li2021}, and trim all clean signals to 6 seconds before convolution. The STFT is calculated using a 480-sample Hamming window, 512-point FFT, and 75\% overlap between successive frames. The features have dimensions of 771$\times$442. 
We evaluate real environments with the ACE challenge \cite{eaton2015ace} and BUT Speech{\textcircled{a}}FIT Reverb corpora {\cite{szoke2019building}}. ACE uses RIRs that were captured in real environments, where the reverberant signals were generated by convolving the RIRs with clean speech. The BUT corpus contains re-transmitted signals from the LibriSpeech corpus \cite{panayotov2015librispeech}. These two corpora will show the generalization capabilities of our approach in unseen and real environments. The ACE dataset (Table \ref{real_rir}) contains seven settings that include longer distances between the microphone and the speaker, where the reverberation times range from 0.34 s to 1.25 s. The BUT corpus contains five room settings with respective dimensions of: 17.2 m $\times$ 22.8 m $\times$ 6.9 m, 4.6 m $\times$ 6.9 m $\times$ 3.1 m, 7.5 m $\times$ 4.6 m $\times$ 3.1 m, 6.2 m $\times$ 2.6 m $\times$ 14.2 m, 10.7 m $\times$ 6.9 m $\times$ 2.6 m, where reverberation times range from 0.61 s to 1.85 s. The average distance between the microphone and the speaker is from 1.41 m to 7.93 m for the BUT corpus, so both the simulated and real environment dataset have signals that differ from 1m. The average distance ranges from 1.35 m to 2.14 m for the ACE corpus. In total we have $1022 + 2620 \times 5 = 14122$ reverberant signals in this testing set.

\subsection{Setup}

In Fig. \ref{fig:2}(B), 16 kernel filters are used for the first two Conv layers, 32 kernel filters for the middle two Conv layers, and 64 kernel filters for the last two Conv layers. All kernel sizes are set to 3 $\times$ 3, with 2 $\times$ 2 kernel sizes for all max pooling layers. The average pooling layer in Fig. \ref{fig:2}(C) uses a kernel size of 3 $\times$ 3. Three fully connected (FC) layers are used  in the $T_{60}$ classification block. The first two FC layers are followed by a BN layer with a leaky ReLU activation function with slopes set to 0.1. The last FC layer also uses ReLU.

The dimensions for the input features to the dereverberation block are 257$\times$442, which is the dimension of the compressed magnitude. It is included with the $T_{60}$ feature vector during the fine tuning stage. For the LSTM in the dereverberation block, the hidden size is set to 512, and the drop out rate is 0.5 to avoid overfitting. We set $\beta$ to 0.9, and $\alpha$ to 0.1 within $L^{A}$ (e.g., Eq. (\ref{cost1})), since these values outperform other options. 

All the pre-trained models use a batch size of 50 and a learning rate of 0.001. However, different optimizers are applied to the two tasks: RMSprop optimizer for $T_{60}$ estimation and Adam optimizer is used for dereverberation. For the joint-learning network, we use the same optimizer and learning rate for the two sub-networks, while the batch size is changed to 64. All the models are trained using the standard backpropagation algorithm for 100 epochs during the pre-training stage and 60 epochs during fine-tuning. We experimented with different $\gamma$ values in the cost function of Eq. (\ref{cost2}). We show results when $\gamma = 0.2$, $\gamma = 1$ (update the weight matrix with only the $T_{60}$ estimation network loss), and when $\gamma = 0.7$ that slightly balances the loss between the two sub-networks. We also provide results from the dereverberation-only network that uses randomly initialized LSTM states.

\section{EVALUATION}
\label{sec:eva}

\begin{table*}[htb]
\caption{Dereverberation: Four Simulated Rooms. $\star$ denotes proposed approach is significantly higher than the four baseline approaches (t-test). ($\cdot$) denotes the standard deviation.}
\centering

\label{tab:2}

{\begin{tabular}{c|ccc|c|ccc|c|ccc|c}
\hline
                             & \multicolumn{4}{c|}{PESQ$\star$} &      \multicolumn{4}{c|}{STOI$\star$} & \multicolumn{4}{c}{SDR} 
                             
                             \\  \cline{2-13}
                             & 0.3           & 0.6                     & 0.9    & AVG        & 0.3     & 0.6       &
                             0.9     & AVG & 0.3 & 0.6 & 0.9 & AVG\\ \hline
\multicolumn{1}{c|}{Unprocessed } & 4.08              &    2.74           &  2.25             & 3.02   & 0.9970   & 0.9381   & 0.8647     & 0.9333 & 31.36 & 12.94 & 8.16 & 17.49\\ \hline
WPE \cite{yoshioka2012generalization,nakatani2010speech}                   & 3.80          &   3.02        &  2.42        & 3.08 (0.31)  & 0.9818    &  0.9502  & 0.8954    &  0.9425 (0.04)  & 28.46 & \textbf{16.71} & 10.58 & 18.58 (4.07)\\ 
LSTM \cite{zhao2018late} & 3.70 & 3.13 & 2.65 & 3.15 (0.33) & 0.9803 & 0.9501 & 0.8999 & 0.9434 (0.03) & 30.51 & 15.21 & 10.07 & 18.60 (4.03) \\ 
RTA \cite{wu2016reverberation} & 4.03 & 3.15 & 2.61 & 3.26 (0.29) & 0.9950 & 0.9503 & 0.8998 & 0.9350 (0.03)& 30.96 & 15.04 & 11.02 & 19.00 (4.26)\\
TeCANet \cite{wang2021tecanet} & 4.11 & 3.22 & 2.69 & 3.34 (0.28) & 0.9973 & 0.9571 & 0.9103 & 0.9549 (0.03)& 31.43 & 15.45 & 11.32 & 19.40 (3.84) \\\hline
Dereverb Only   & 4.20          & 3.19 & 2.66 & 3.35 (0.31) & 0.9972   & 0.9553   & 0.9062   & 0.9529 (0.03) & 31.04 & 15.13 & 11.10 & 19.09 (4.64)       \\ 
Proposed ($\gamma$=0.2)                         & \textbf{4.26}         & 3.44          & 2.91        & 3.54 (0.27)  & \textbf{0.9975}   & 0.9644   & 0.9276    & 0.9631 (0.02) & \textbf{31.88} & 16.31 & 12.49 & 20.23 (3.71)    \\ 
Proposed ($\gamma$=0.7)                         & 4.25         & \textbf{3.47}          & \textbf{2.97}        & \textbf{3.56} (0.27)   & 0.9974   & \textbf{0.9650}   & \textbf{0.9304}    & \textbf{0.9643} (0.02) & 31.80 & 16.37 & \textbf{12.63} & \textbf{20.27} (3.69)    \\
Proposed ($\gamma$=1)                         & 4.24         & 3.45          & \textbf{2.97}       & 3.55 (0.27)  & 0.9974   & 0.9644   & 0.9295    & 0.9638 (0.02)& 31.75    & 16.27 & 12.55 & 20.19 (3.70)     \\
\hline

\end{tabular}}

\end{table*}

\begin{table*}[ht!]
\centering
\caption{Performance comparison of the proposed approach for $T_{60}$ estimation in the four simulated rooms}
\label{tab:4}
{\begin{tabular}{c|cc|cc|cc|cc}
\hline
                             & \multicolumn{2}{c|}{MSE} &      \multicolumn{2}{c|}{MAE} & \multicolumn{2}{c|}{$\rho$} &
                             \multicolumn{2}{c}{$\eta$}
                             \\  \cline{2-9}
                             & Reg           & Cls                     & Reg    & Cls        & Reg     & Cls       &
                             Reg     & Cls\\ \hline
\multicolumn{1}{c|}{MLP \cite{mlp}} & 0.253               &   \(-\)            &  0.334              & \(-\)    &  0.683   &  \(-\)  &  0.703     &  \(-\)   \\ 
CNN \cite{bryan2020impulse}                   &  0.271          &  \(-\)         &  0.291         & \(-\)   &  0.792   &  \(-\)  &  0.824     &    \(-\)   \\ \hline
$L_{joint}^B(\gamma=0.2)$                        & 1.544 & 0.032          & 1.219          & 0.121   & 0.459   & 0.844   & 0.032    & 0.852    \\
$L_{joint}^B(\gamma=0.7)$                        & 0.903 & 0.035          & 0.938          & 0.126   & \textbf{0.889}   & 0.799   & \textbf{0.905}    & 0.751    \\
$L_{joint}^B(\gamma=1)$                        & 0.417 & \textbf{0.021}          & 0.613          & \textbf{0.094}   & 0.602   & 0.844   & 0.631    & 0.840    
\\\hline
\end{tabular}}

\end{table*}

\subsection{Baseline Approaches and Evaluation Metrics}

We evaluated our proposed approach with two main tasks: (1) dereverberation and (2) $T_{60}$ estimation. The following first provides details of the baseline approaches for the dereverberation task and then for the $T_{60}$ estimation task.

We implemented five baseline comparison approaches. The first approach uses a statistical model-based approach, known as weighted prediction error (WPE) that estimates an inverse filter to remove the late reverberation from a given reverberant signal \cite{yoshioka2012generalization,nakatani2010speech}. The second approach uses a LSTM to estimate late reverberation \cite{zhao2018late} and subtracts that from the original signal to produce the direct-early component. The third approach uses a reverberation time-aware (RTA) DNN \cite{wu2016reverberation} and a context window to directly estimate the magnitude of the anechoic speech. Similarly, the fourth approach \cite{wang2021tecanet}, called TeCANet, uses a context aware input and a Full-Band based Temporal Attention (FTA) approach, where the current frame is the key and context frames are the query and value. TeCANet directly predicts the anechoic magnitude and reconstructs the waveform using the reverberant phase. We modify RTA and TeCANet so that they predict the direct-early component, in order to make all frameworks consistent for comparison. We lastly compare with a dereverb only model from section \ref{early}, where this model differs from our proposed model by only providing magnitude information as input and training the model to learn the magnitude of the direct-early component without knowing any room acoustic information.

We evaluate dereverberation performance using the perceptual evaluation of speech quality (PESQ) \cite{pesq}, short-time objective intelligibility (STOI) \cite{stoi}, and the signal to distortion ratio (SDR) \cite{bss_eval}. The reference signal, in each case, is the direct-early component of the reverberant signal. 

We implement two $T_{60}$ estimation comparison approaches: (1) CNN \cite{bryan2020impulse} and (2) spectro-temporal modulation filtering and a MLP \cite{mlp}. Note that we did not apply data augmentation when training the CNN. We use the log-mel spectrogram as the input to the CNN, and the exact architecture as in \cite{bryan2020impulse}. For the MLP, we use the Gabor 2D filters to extract the input, and a 3-layer MLP for classification \cite{mlp}.

We compute the mean-square error (MSE), mean-absolute error (MAE), Pearson's correlation coefficient (PCC, $\rho$) and Spearman's Rank Correlation Coefficient (SRCC, $\eta$) between the ground truth and the estimated $T_{60}$. For MSE and MAE, smaller values indicate better performance, whereas scores closer to 1 are better for PCC and SRCC.
\subsection{Results for Simulated Data}
\label{sim_res}

\begin{figure*}[!htb]
\minipage{0.33\textwidth}
  \includegraphics[width=\linewidth]{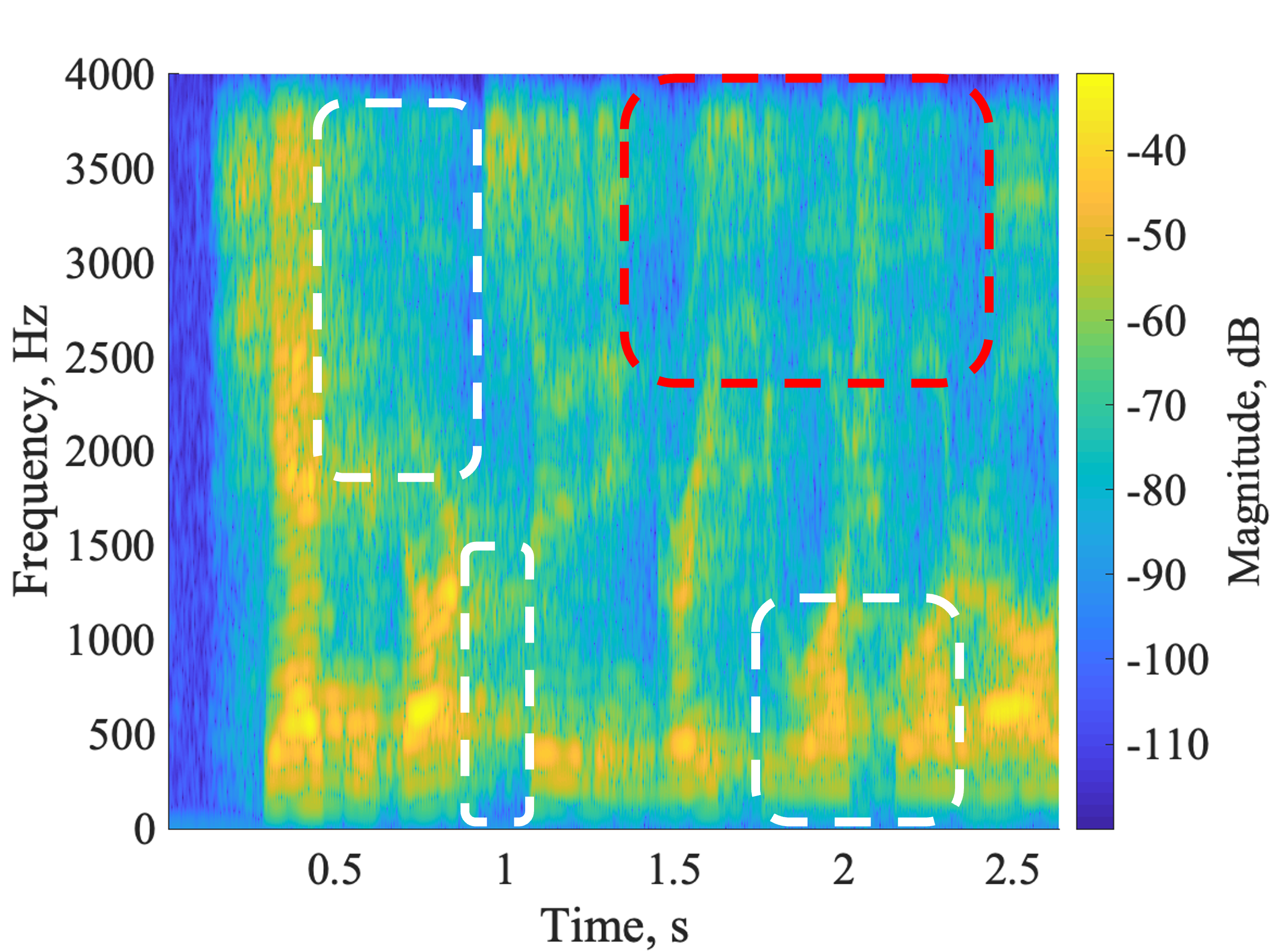}
\endminipage\hfill
\minipage{0.33\textwidth}
  \includegraphics[width=\linewidth]{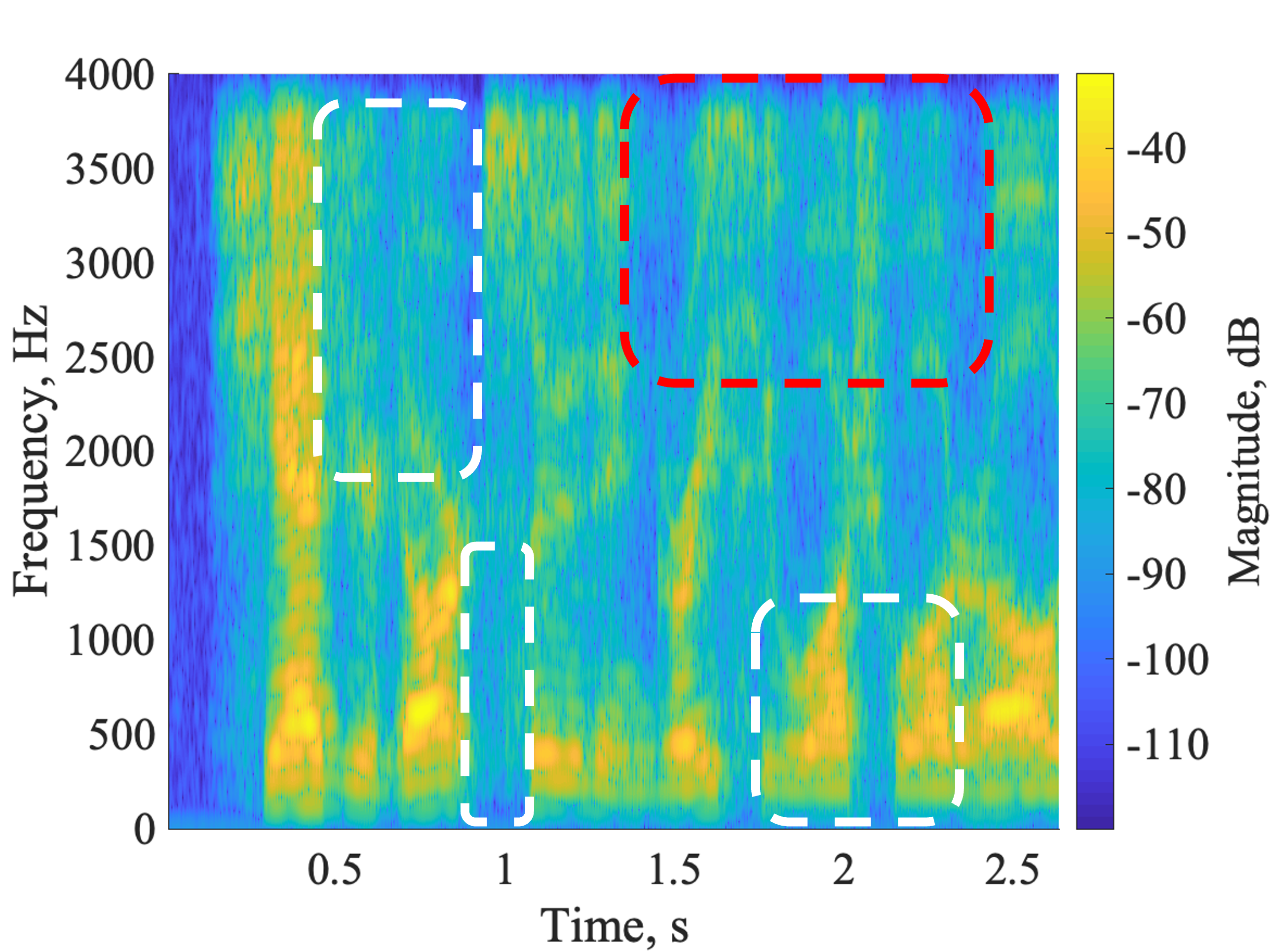}
\endminipage\hfill
\minipage{0.33\textwidth}%
  \includegraphics[width=\linewidth]{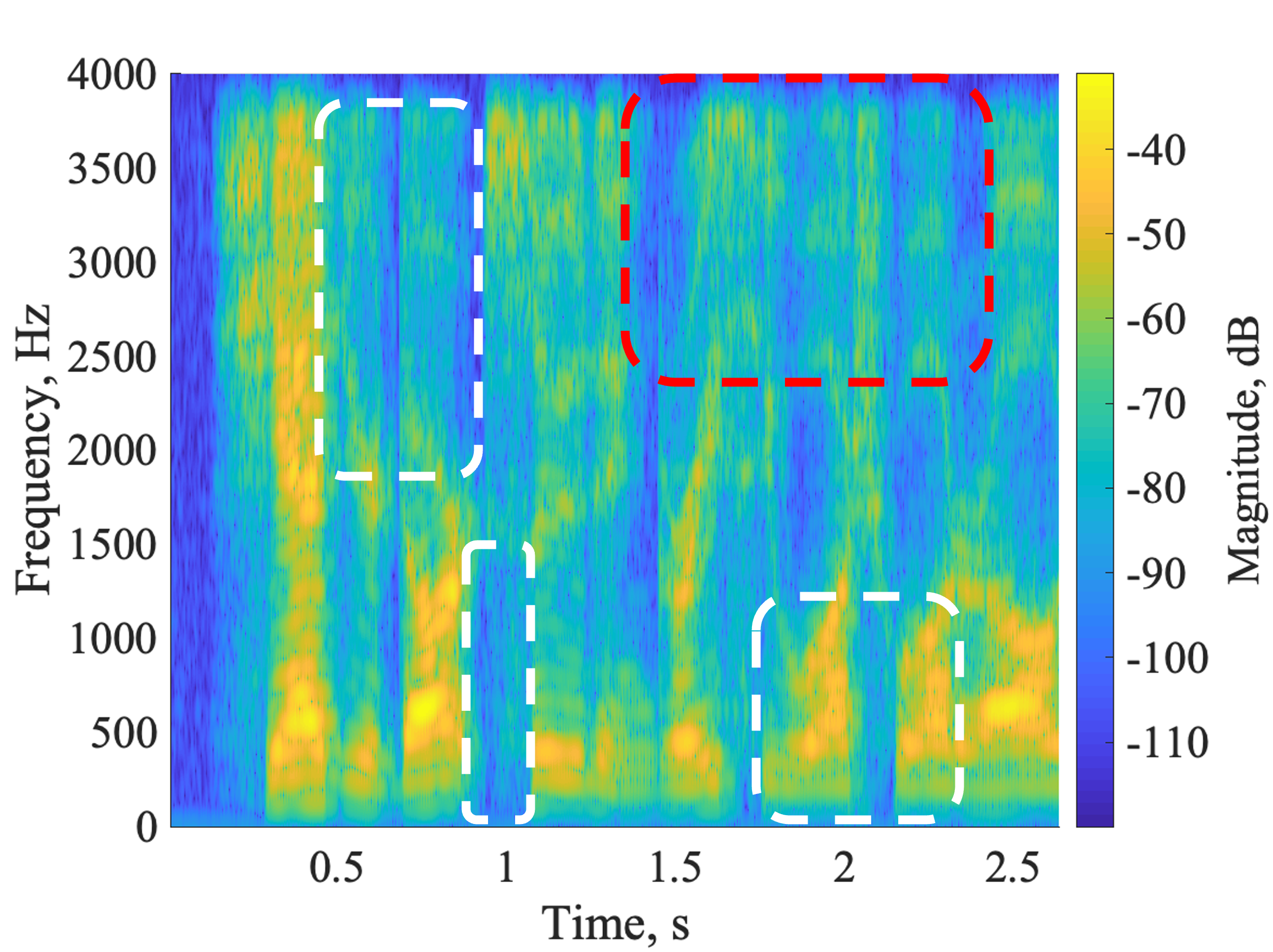}
\endminipage
\caption{Spectrograms for (left) the simulated reverberant signal ($T_{60}=0.9\, s$), (middle) the estimated direct-early signal, and (right) the ground truth direct-early signal.}
\label{fig:spectro}
\end{figure*}

The results in Table \ref{tab:2} show that the overall performance across the three metrics for our proposed approach is better than all other comparison approaches, including the dereverb only model, in simulated environments. When $T_{60} = 0.3\, s$, there is less reverberation, and it sounds almost clean to human ears, which is evidenced by the high PESQ score (4.08) for the unprocessed reverberant signal. Our proposed approach produces a gain of 0.18 for PESQ, a negligible STOI improvement, and an overall SDR improvement of at least 0.52 dB, which is the largest gain. Most of the comparison approaches failed to perform dereverberation, where TeCANet generated a 0.07 PESQ gain. 

With mild reverberation ($T_{60} = 0.6\, s$), the proposed approach improves PESQ by 0.73, while the best score from comparison approaches improves PESQ by 0.48. The proposed approach also provides the largest gain in STOI, which is approximately 0.027. However, for SDR, WPE outperforms all approaches with an approximate gain of 3.77, which is only slightly higher than our proposed approach. A likely reason that WPE performs better for this $T_{60}$ is that unsupervised learning predicts the transition time that is appropriate at each $T_{60}$, while the other approaches assume a fixed transition time across all conditions, and are optimized across the average. Without the appropriate mixing delay, the models may introduce more distortions that lower the SDR score.

At the longest reverberation time ($T_{60} = 0.9\, s$), the proposed approach produces the largest improvement according to all three metrics, with improvement scores of 0.72, 0.0657 and 4.47 for PESQ, STOI and SDR, respectively. Overall, when averaging scores across the three $T_{60}s$, the proposed system outperforms all the comparison approaches, including the dereverb only model, which provides evidence that additional $T_{60}$ information helps improve the dereverberation performance. We provide results using less weight on the $T_{60}$ estimation loss ($\gamma = 0.2$), and heavier weight on the $T_{60}$ estimation loss ($\gamma = 0.7$) and only $T_{60}$ estimation loss ($\gamma = 1$) during the finetuning stage of our proposed method. The average score for PESQ and STOI are quite stable even when $\gamma = 1$, and the average score of SDR shows a subtle difference (0.14) when comparing $\gamma = 0.2, \, 0.7$ and $1$.  On average, our proposed approach provides the lowest standard deviation for every evaluation metric, and all three scores are (statistically) significantly better than all four baseline approaches ($p < 0.05$, t-test) as well as the dereverb only model.

Table \ref{nonshoe} shows the evaluation results for non-rectangular rooms to further evaluate model generalization. We compare our proposed model with WPE and the baseline approach with highest performance in other simulation conditions, TeCANet. Our model produces the largest gain with respect to PESQ, STOI and SDR, where the respective improvements are 0.43, 0.111 and 2.22. 

Table 
\ref{tab:4} shows the evaluation results for $T_{60}$ estimation after the fine-tuning stage. We place more weight on the classification-based regression subtask (e.g., $\alpha = 0.9$ in Eq. (\ref{cost1})), and the overall scores reflect that the jointly-trained model performs better on classification-based estimation, and especially for MSE and MAE. We compare our proposed system with two baseline approaches. Bryan \textit{et al.} \cite{bryan2020impulse} uses a CNN to estimate the direct-to-reverberant ratio (DRR) and \(T_{60}\). Xiong \textit{et al.} \cite{mlp} estimate \(T_{60}\) based on spectro-temporal modulation filtering and a DNN. The results show that our model outperforms the baseline approaches. 
The classification branch where $\gamma = 1$ gives the best MSE and MAE results, however,  $\gamma = 0.7$ gives the best performance in PCC and SRCC with the regression branch. 
The MSE and MAE results are better in terms of classification-based regression, but produce worse scores in terms of PCC and SRCC. This likely happens because during joint learning, we only use the classification branch with the MSE and cross-entropy loss terms. Therefore, the proposed approach minimizes the MSE and MAE, but this may negatively impact PCC and SRCC.

\begin{table}[htb]
\caption{Performance comparison of the proposed approach for the dereverberation task in the non-rectangular simulated rooms.}
\centering

\label{nonshoe}

{\begin{tabular}{c|c|c|c}
\hline
                             & \multicolumn{1}{c|}{PESQ} &      \multicolumn{1}{c|}{STOI} & \multicolumn{1}{c}{SDR}
                             
                             \\  \cline{1-4}
\multicolumn{1}{c|}{Unprocessed } & 2.33              &    0.8790           &  8.94           \\ \hline
WPE \cite{yoshioka2012generalization,nakatani2010speech}                   & 2.57          &   0.9095        &  10.92\\ 
TeCANet \cite{wang2021tecanet} & 2.63 & 0.9096 & 11.03  \\\hline
Proposed ($\gamma$=0.2)                         & 2.75         & 0.9096          & \textbf{11.16}          \\ 
Proposed ($\gamma$=0.7)                         & \textbf{2.76}         & \textbf{0.9099}          & \textbf{11.16}          \\
Proposed ($\gamma$=1)                         & \textbf{2.76}         & \textbf{0.9099}          & 11.13     \\
\hline

\end{tabular}}

\end{table}

Fig. \ref{fig:spectro} shows the spectrograms of a reverberant signal, an estimated output signal from our proposed approach, and the ground truth of the direct-early reference signal. The magnitude of the reverberant signal is quite blurry and distorted, especially in the high frequency band that is smeared by the reflections (particularly, the late reflections). This is evidenced by the difficulty in seeing the silent periods between each word. Fig. \ref{fig:spectro} (middle) shows the estimated direct-early component from our proposed system, which is highly similar to the ground truth (right), where reverberation is clearly removed in between the words (see the white bounding box in Fig. \ref{fig:spectro} (middle)). However, some of the high frequencies are still smeared by late reflections (see the red bounding boxes).

\begin{table*}[htb]
\centering
\caption{Performance comparison of the proposed approach in terms of PESQ using the real reverberant ACE corpus. \(-\) means the same as in direct sound}
\label{real_pesq}

\scalebox{0.86}
{\begin{tabular}{c|ccccccc|c|ccccccc|c}
\hline
                             & \multicolumn{8}{c|}{Direct sound} 
                    &\multicolumn{8}{c}{Direct-early component}
                             
                             \\  \cline{2-17}
                             & Lobby           & Meet 1                     & Meet 2 & Lec 1 & Lec 2 & Of 1 & Of 2    & AVG   & Lobby           & Meet 1                     & Meet 2 & Lec 1 & Lec 2 & Of 1 & Of 2    & AVG     \\ \hline
\multicolumn{1}{c|}{Unprocessed } & 2.11  &2.06 &2.12 &2.35 &  1.91 & 2.10 & 2.12 & 2.11   &\(-\)  &\(-\) &\(-\) &\(-\) &\(-\) &\(-\) &\(-\)  &\(-\)             \\ 
WPE \cite{nakatani2010speech}                  &\(-\)  &\(-\) &\(-\) &\(-\) &\(-\) &\(-\) &\(-\)  &\(-\)   & 2.29 & 2.21 & 2.27 & 2.69 & 2.04 & 2.15 & 2.29 & 2.28                 \\ 
LSTM \cite{zhao2018late} & 2.32 &2.19 & 2.20 & 2.65 & 2.12 & 2.17 & 2.25 & 2.27 &2.12 &2.15 &2.14 &2.57 &1.98 &2.15 &2.19 &2.19 \\
RTA \cite{wu2016reverberation}                   & 2.44 & 2.24 & 2.35 & 2.70 & 2.26 & 2.28 & 2.34 & 2.37  &2.35 &2.20 &2.29 &2.67 &2.13 &2.17 &2.20 &2.29       \\ 
TeCANet \cite{wang2021tecanet} & 2.49 & 2.25 & 2.37 & 2.73 & 2.30 & 2.31 & 2.36 & 2.40 &2.39 &\textbf{2.22} &\textbf{2.32} &\textbf{2.69} &2.20 &2.20 &2.24 &2.32 \\ \hline
Dereverb Only & 2.21 & 2.10 & 2.19 & 2.45 & 1.97 & 2.15 & 2.18 & 2.18 &2.19 &2.10 & 2.17 & 2.41 &1.96 &2.15 & 2.19 & 2.17\\
Proposed ($\gamma$=0.2)                         & 2.55 & \textbf{2.27} & \textbf{2.43} & \textbf{2.80} & \textbf{2.34} & \textbf{2.35} & \textbf{2.42} & \textbf{2.45}  &2.41 &2.20 & 2.31 &\textbf{2.69} &2.25 &\textbf{2.23} &2.28 &\textbf{2.34}  \\
Proposed ($\gamma$=0.7)                         & \textbf{2.56} & 2.26 & \textbf{2.43} & \textbf{2.80} & \textbf{2.34} & 2.34 & 2.41 & \textbf{2.45} &\textbf{2.42} &2.20 &2.30 &2.68 &\textbf{2.26} &2.22 &\textbf{2.29} &\textbf{2.34}   \\
Proposed ($\gamma$=1)                         & 2.55 & 2.25 & 2.42 & 2.78 & 2.30 & 2.33 & 2.40 & 2.43  &2.41 &2.20 &2.30 &2.67 &\textbf{2.26} &2.22 &2.28 &2.33  \\\hline
Proposed (finetune) & 2.50 & $\cdot$ & \textbf{2.57} & $\cdot$ & $\cdot$ & $\cdot$ & 2.55 &$\cdot$ & \textbf{2.51} & $\cdot$ & \textbf{2.56} & $\cdot$ & $\cdot$ & $\cdot$ & \textbf{2.45} &$\cdot$ \\\hline

\end{tabular}}

\end{table*}

\begin{table}[htb]
\caption{Performance comparison of the proposed approach (direct sound estimation) for the dereverberation task applied to real reverberant speech from the BUT Speech{\textcircled{a}}FIT Corpus}
\centering

\label{re_BUT}

{\begin{tabular}{c|c|c|c}
\hline
                             &      \multicolumn{1}{c|}{PESQ} & \multicolumn{1}{c|}{STOI} &\multicolumn{1}{c}{SDR}
                             
                             \\  \cline{1-4}
\multicolumn{1}{c|}{Unprocessed }               &    1.48           &  0.2735       & -5.11    \\ 
WPE \cite{yoshioka2012generalization,nakatani2010speech}                             &   \textbf{1.50}        &  \textbf{0.2757} & -3.93\\ \hline
Proposed ($\gamma$=0.2)                                 & 1.48          & 0.2736        & \textbf{-3.45}  \\ 
Proposed ($\gamma$=0.7)                         & 1.47         & 0.2712         &-3.76        \\
Proposed ($\gamma$=1)                                 & 1.48          & 0.2701   &-6.86  \\
\hline

\end{tabular}}

\end{table}
\begin{table}[htb]
\centering
\caption{Performance comparison of the proposed approach (direct sound estimation) in terms of DNSMOS on the real reverberant speech of the ACE corpus}
\label{real_DNSMOS}

\scalebox{0.86}

{\begin{tabular}{c|ccc}
\hline
                             & \multicolumn{3}{c}{Direct sound} 
                             
                             \\  \cline{2-4}
                             & Lobby                            
                             & Meet 2 
                             & Of 2         
                             \\ \hline
\multicolumn{1}{c|}{Unprocessed } & 2.94   
&3.03 
& 2.96             
\\ 
WPE \cite{nakatani2010speech} & 3.02 & 3.11 & 3.09 \\
TeCANet \cite{wang2021tecanet} & 2.90 
& 3.05 
& 3.16 
\\ \hline
Proposed ($\gamma$=0.2)                         & 3.02 
& 3.10
& 3.12 
\\
Proposed ($\gamma$=0.7)                         & 3.03 
& 3.11 
& 3.11 
\\
Proposed ($\gamma$=1)                         & \textbf{3.04} 
& \textbf{3.16} 
& \textbf{3.22} 
\\\hline

\end{tabular}}

\end{table}

\subsection{Results for Real Data}
\label{real_res}

Although our approach produces improvements in the simulated testing environments, we still need to perform tests using real RIRs and signals to determine if our proposed approach is robust and generalizes to real environments. As discussed in section \ref{sec:exp}.A we evaluate real performance using the ACE challenge dataset \cite{eaton2015ace} and the BUT Speech@FIT reverberation corpus \cite{szoke2019building}. 
The ACE challenge dataset has the clean speech signal instead of the direct-early component as the reference. This means that the degraded signals are compared to the clean signal as opposed to the direct-early signal, which will result in lower scores. Additionally, each real room has different objects, which may also result in lower PESQ scores. 

For a more precise comparison, we ran two experiments: (1) predicting the direct-early component as in section \ref{joint} and (2) modifying our training target to the direct sound, and retraining our model with simulated training set. For the other deep learning based comparison approaches, we retrain the model for the direct sound target. It is worth noting that the direct sound has a time delay compared with the clean signal, so the reference signal is not time aligned, which will lower certain scores (e.g., STOI and SDR). We are unable to modify WPE that estimates the direct-early component, since it is a unsupervised approach. We use the same network architecture from section \ref{sim_res} and train using 15000 simulated signals with three T$_{60}$s: 0.3 s, 0.6 s and 0.9 s. 

Table \ref{real_pesq} shows the average PESQ scores when predicting the direct sound and the direct-early component for each room when using the real RIRs from the ACE dataset. 
For the direct-early component estimation, our proposed approach shows the greatest gain for five out of seven rooms. The average PESQ score improvement of our proposed system outperforms all other approaches. 
For direct sound estimation, the overall scores for deep learning based approaches are better. Specifically, our proposed approach, WPE, LSTM RTA and TeCANet show PESQ improvements, but our proposed approach shows the greatest gain for every room (e.g. 0.34 in Office 1) amongst all the approaches. Additionally, we randomly selected four out of the seven rooms in the ACE corpus for fine-tuning with real data. We then evaluate performance with the remaining three rooms. The last row shows the PESQ scores for the three testing rooms after fine-tuning. The results show significant PESQ improvement compared to previous model in both direct sound estimation and direct-early component estimation. WPE and our proposed approach performed best and nearly identical in terms of STOI (0.83) and SDR (12 dB).

Table \ref{re_BUT} shows the average scores when predicting the direct sound using the BUT Speech\textcircled{a}FIT Database after fine-tuning with real data. We fine-tuned the model by randomly selecting three of the five rooms for training, and the remaining two for testing. Our proposed approach performed best on average SDR score with the largest gain 1.66. WPE performed best according to PESQ and STOI, but these results are comparable to ones from our proposed approach. Note that the overall scores are lower for this corpus, which speaks to the difficulty of removing reverberation in real environments.

Additionally, we use a multi-stage data-driven perceptual objective metric known as the Deep noise Suppression Mean Opinion Score (DNSMOS) to evaluate our proposed dereverberation estimation approach for 
real reverberant speech in 
Table {\ref{real_DNSMOS}}. DNSMOS uses machine learning to estimate human-evaluated MOS {\cite{reddy2021dnsmos}.

The results show the DNSMOS score for the ACE corpus after the fine-tuning stage with real environment data. Our proposed approach shows the greatest gain of 0.1, 0.13 and 0.26 with respective of Lobby, Meet 2 and Office 2. This indicates that our approach is also best according to human evaluators.}

\section{Discussion}
\label{sec:con}

\subsection{Results on different weight parameter $\gamma$}

From Table 
\ref{tab:2} to \ref{tab:4}, we provide proposed approach results for different $\gamma$ values in simulated unseen rooms for the dereverberation and $T_{60}$ estimation tasks. For the dereverberation task, when $\gamma = 0.7$, the proposed approach performs best for both PESQ and STOI scores, and when $\gamma = 0.2$, the proposed approach performs the best on SDR when $T_{60} = 0.3\, s$. However, the overall results from different $\gamma$ values did not show much difference between each other, 
where the largest gain is 0.01 for PESQ, 0.09 for SDR when $T_{60} = 0.3\, s$, and 0.0029 for STOI when $T_{60} = 0.9\, s$. These slight differences according to the three evaluation metrics indicate that $T_{60}$ features are more crucial to improving dereverberation performance, than the fine-tuning stage regardless of the $T_{60}$ estimation loss included during the back propagation process. 

Different patterns are found for the $T_{60}$ evaluation scores in Tables 
\ref{tab:4}. For the pure-regression task, the scores change dramatically for different $\gamma$ values across the four different metrics. However, classification-based regression benefits from joint learning due to parameter sharing for the related tasks of room acoustic estimation and dereverberation. This is especially evident in Table \ref{tab:4}, where our proposed approach provides the best scores overall and shows the ability to generalize. The results in Table 
\ref{tab:4} also explain why we we use the classification branch instead of the regression branch for the joint-learning connection, since the results show that the classification branch's outputs are more reliable. 
 
\subsection{Incorporating $T_{60}$ information}

Section \ref{early} describes our direct-early component estimation (dereverb only) model. The idea is similar to LSTM \cite{zhao2018late}, where we randomly initialize the hidden and cell states. This dereverb only model does not have prior information of the room characteristics, and the results from Table 
\ref{tab:2} show  that it is more difficult to perform dereverberation as $T_{60}$ increases, without this room environment information. One way to incorporate the room characteristics is to provide $T_{60}$ information during the training stage (RTA) \cite{wu2016reverberation}, where a pair of parameters (e.g., frame shift and context window size) are provided as additional inputs. Theses parameters influence the STFT calculation for the input signal and the resulting feature size. During training, the network is provided with different input features based on the $T_{60}$, where the extra information allows the network to more accurately learn the anechoic speech. As for the testing stage, this method requires an external $T_{60}$ estimator and a lookup table to determine the corresponding frame shift and context window size. This approach shows that $T_{60}$ is necessary and even helpful for dereverberation, however, the external $T_{60}$ estimator is not part of the overall network, which could be sub-optimal. 

In order to find the optimal $T_{60}$ information, we also investigated other experiments in which we pass the classification results or the $T_{60}$ estimation results as inputs to the dereverb module. Those results are not comparable to RTA, which shows that this information is not as helpful to dereverberation performance. Our proposed approach, on the other hand, addresses this by learning important $T_{60}$ features and providing them to the dereverberation network through a skip connection. By comparing the results from RTA, our proposed approach, and the dereverb only model in Table \ref{tab:2}, the three approaches respectively produce average PESQ improvements of 0.24, 0.54 and 0.33, STOI improvements of 0.0017, 0.031 and 0.0191, and SDR improvements of 1.51, 2.78 and 1.6 dB over the unprocessed signal. The results indicate that the prior information provides different contextual information and an alternative input feature that is beneficial to performance. We surmise that this occurs because the features from the reverberation time estimation help distinguish between different levels of reverberation. In Fig. \ref{tsne}, we use t-SNE to visualize the output of the penultimate layer from the $T_{60}$ classification block. The $T_{60}$ features from the pre-trained module are clustered together according to reverberation time, which could be a key identifier for the direct-early component estimation module. t-SNE visualizations provide similar results when used for other self-supervised algorithms \cite{chi2021audio}. 
\begin{figure}
\begin{center}
\includegraphics[width=\columnwidth,height=6cm]{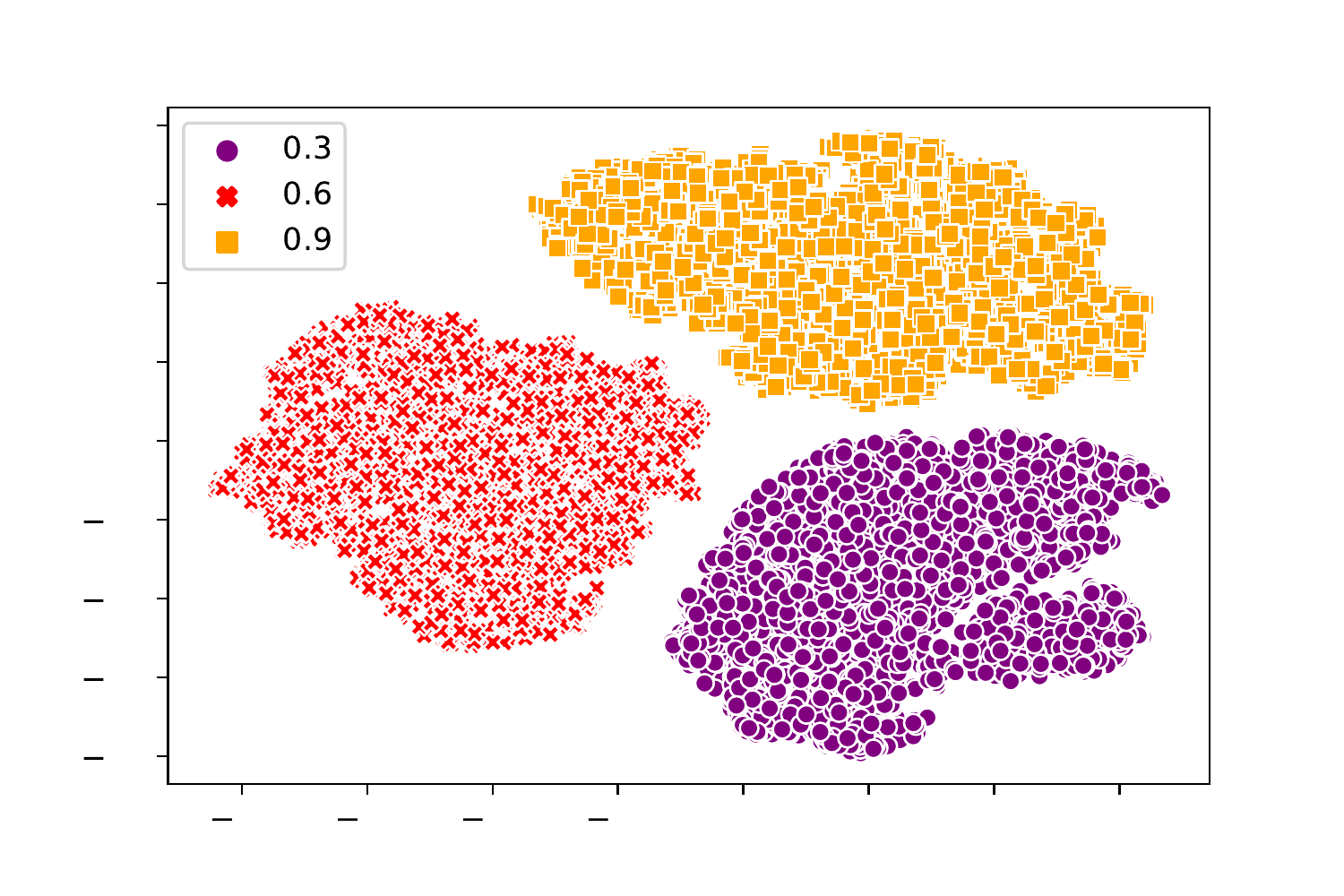}
  \caption{(Color Online) Visualization of the $T_{60}$ classification feature representations using t-SNE on the simulated training data.}
\vspace{-2em}
  \label{tsne}
\end{center}
\end{figure}

\subsection{Limitations and future work}

Section \ref{sec:eva} discussed the performance of the results based on the proposed model that trained with simulated data. Although the performance was significantly higher than the baseline approaches ($p < 0.05$, t-test) on simulated test data, one limitation that needs to be addressed is appropriately leveraging real environment data. With simulated data, we can generate a balanced and large dataset to meet requirements for deep learning. With real data, however, we generally do not have enough data since recordings often do not have the corresponding ground truth signal or acoustic parameters that are needed for a supervised approach. Furthermore, our proposed approach is based on how well the $T_{60}$ module produces valid $T_{60}$ features. However, for real environments, the $T_{60}$s are often unseen, which negatively impacts $T_{60}$ class prediction.

Our proposed approach provides $T_{60}$ features to the dereverb module for further accurate dereverberation. However, the reverberant training data is generated with the image source method, which finds the path length and pressures of purely specular reflections and has limitations on modelling diffuse reflections that could occur in late reflections. Based on this property of simulated data, our model has been limited to learn the RIR information with inaccurate fuse reflections, which negatively impacts real-world testing.

Instead of estimating $T_{60}$ directly from reverberant speech itself, we could address the dereverberation problem with visual-only data {\cite{alawadh2022room}}, {\cite{kim2020acoustic}} or audio-video data {\cite{li2018scene}}, {\cite{remaggi2018audio}}. For a 3D video, the video will contain the corresponding room environment characteristics, such as room dimensions, the positions of microphone and the speaker, and objects in the room that cause the reflections. This information could provide key indicators of the acoustic characteristics. With accurate room information, we have a better chance to estimate the $T_{60}$, and our proposed model could benefit from this with carefully designed embedded features.  
 
\section{Conclusion}
\label{sec:newcon}
In conclusion, we propose a joint-learning network that integrates $T_{60}$ estimation information into a dereverberation module to enhance reverberant signals. In particular, we provide the penultimate output of the $T_{60}$ estimation module, which serves as a reverberation time feature, along with the compressed magnitude to the dereverberation network. Most importantly, the results show significant improvements in both objective speech quality and intelligibility when providing the $T_{60}$ features to the joint network, where we produce gains in simulated and real environments.

\ifCLASSOPTIONcaptionsoff
  \newpage
\fi

\bibliographystyle{IEEEtran}
\bibliography{strings}

\begin{IEEEbiography}[{\includegraphics[width=1in,height=1.25in,clip,keepaspectratio]{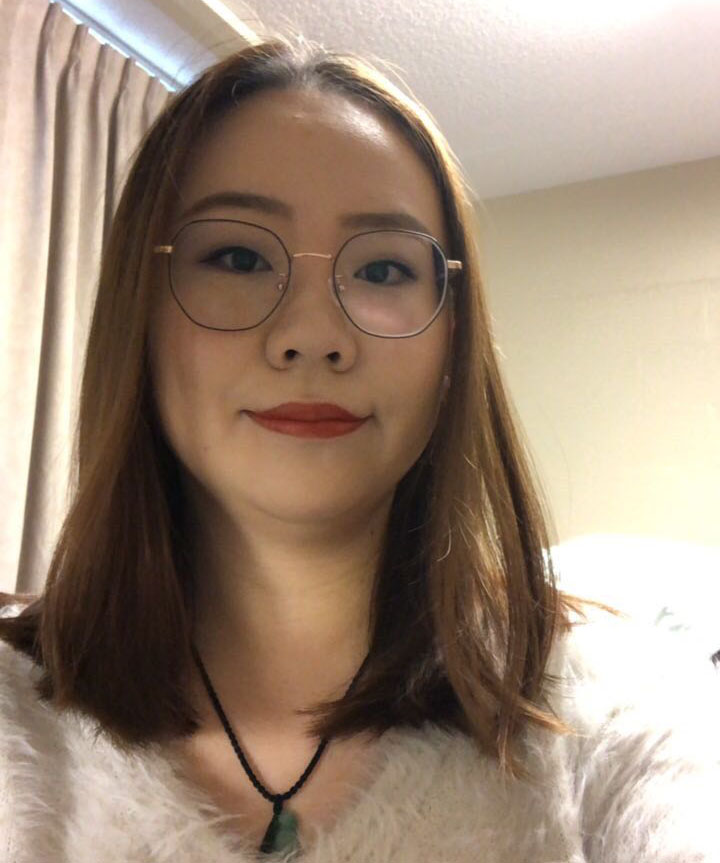}}]{Yuying Li}
received the B.S. degree in computer science and B.S.B. degree in information \& process management from Indiana University, Bloomington, IN, in 2015 and the M.S. degree in computer science from Indiana University, Bloomington, IN, in 2016. She is currently pursuing the Ph.D. degree in intelligent systems engineering at Indiana University, Bloomington. Her research interest include speech dereverberation, machine learning, and audio-visual processing.
\end{IEEEbiography}

\begin{IEEEbiography}[{\includegraphics[width=1in,height=1.25in,clip,keepaspectratio]{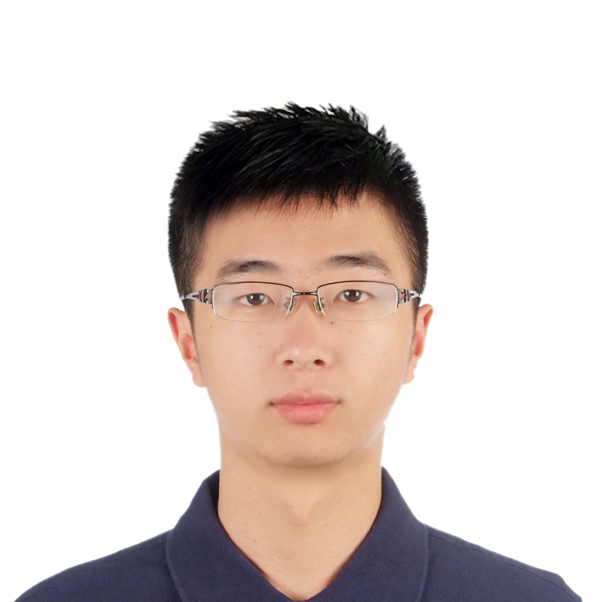}}]{Yuchen Liu}
received the B.S. degree in computer science and financial economics from Centre College, Danville, KY, in 2016 and the M.S. degree in data science from Indiana University, Bloomington, IN, in 2018. He is currently pursuing the Ph.D. degree in computer science at Indiana University, Bloomington. His research interest focus on a variety of speech and audio related deep learning tasks including audio representation learning, audio privacy, speech recognition, and speech assessment.

\end{IEEEbiography}

\begin{IEEEbiography}[{\includegraphics[width=1in,height=1.25in,clip,keepaspectratio]{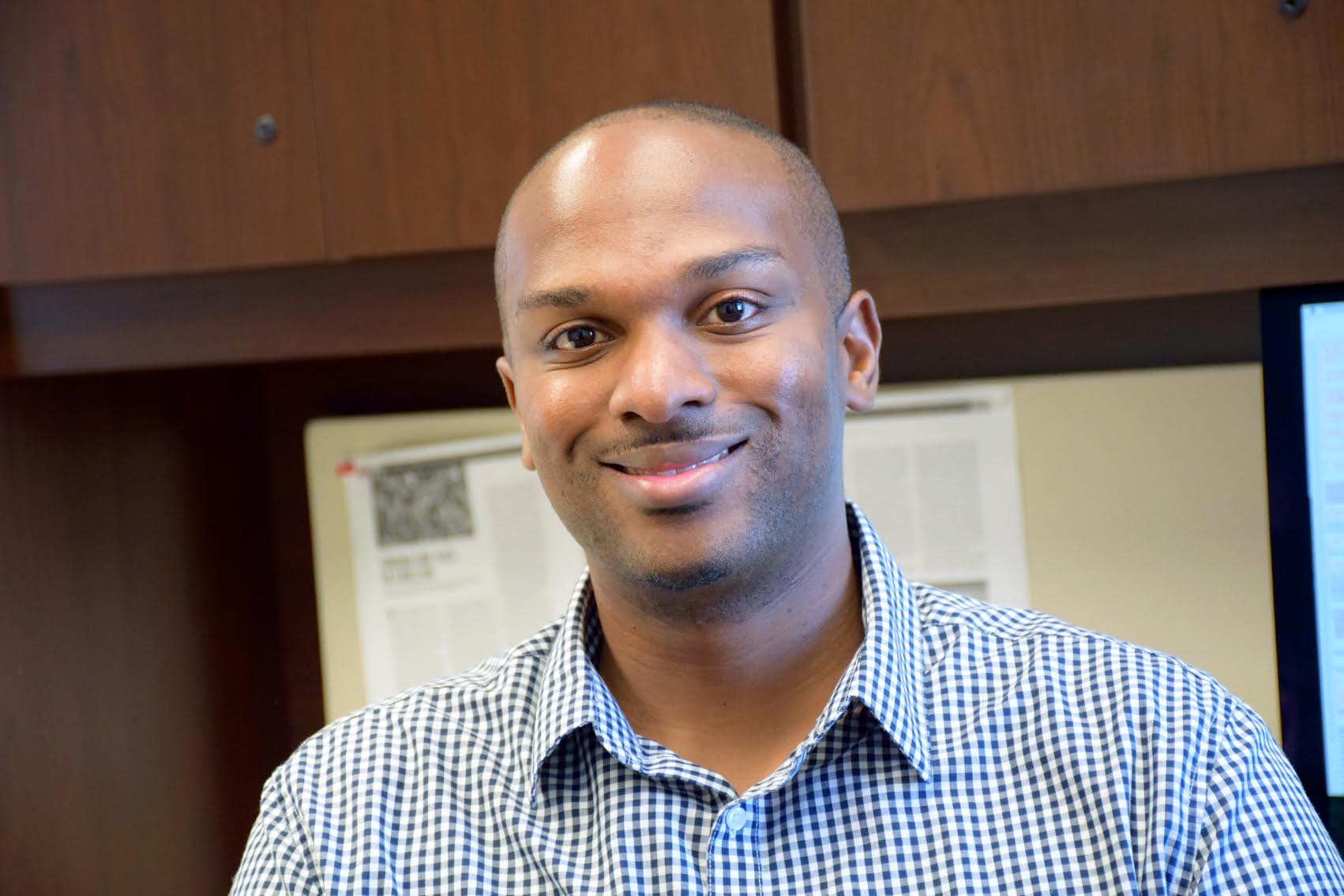}}]{Donald S. Williamson} is currently an Associate Professor with the Department of Computer Science and Engineering at The Ohio State University, USA. He was previously an Assistant/Associate Professor at Indiana University in the Department of Computer Science within the Luddy School of Informatics, Computing and Engineering. His research interests include speech enhancement, speech assessment, and speech privacy.

\end{IEEEbiography}

\vfill

\end{document}